\documentclass[preprintnumbers,twocolumn,amsmath,amssymb]{revtex4}
\usepackage{graphicx}
\usepackage{dcolumn}
\usepackage{bm}
\begin{document}
\title{ The influence of boundaries on high pressure melting experiments}
\author{ V. Sorkin }
\email{phsorkin@techunix.technion.ac.il }
\homepage{ http://phycomp.technion.ac.il/~phsorkin/index.html}
\author{ E. Polturak}
\author{ Joan Adler}
\affiliation{ Physics Department, Technion - Israel Institute of Technology, Haifa, Israel, 32000  }

\date{\today}

\begin{abstract}
At low  pressure, free surfaces play a crucial role in the melting transition.
Under pressure, the surface of the sample is acted upon by some pressure transmitting medium.
To examine the effect of this medium on
melting, we performed Monte Carlo simulations of a system of argon atoms
in the form of a slab with two boundaries.
We examined two cases, one with a soft and the other with a rigid medium at the boundaries.
We found that in the presence of a rigid medium, melting resembles
the mechanical lattice instability found in a surface-free solid. 
With a soft medium at the boundary, melting begins at the surface and at a lower temperature.
The relevance of these results to experiment is discussed.

\end{abstract}

\maketitle

Phase diagrams of materials at  high pressures and temperatures
are of great interest due to their importance for geology and astrophysics,
in particular understanding the Earth's core \cite{Boehler0}.
For example, the melting line of iron under high pressure and temperature
determines the locus of the 
solid-liquid interface inside the Earth's core.
The melting line of rare-gas solids is important for understanding
the abundance of these gases in the  Earth's atmosphere~\cite{Jeff}.

Experimental studies of melting at high pressures 
are performed using the diamond anvil cell (DAC)\cite{Jayaraman}
or the shock wave \cite{Duvall} technique. 
An ongoing controversy exists regarding the melting line of iron~\cite{Kerr,Errandonea}
obtained by these two methods. It seems that the  melting temperature, $T_m$,
determined using shock waves is systematically higher than that measured
in the DAC experiments. This difference introduces a considerable uncertainty
into the models of the Earth's core~\cite{Kerr}.
In addition, Errandonea~\cite{Errandonea} pointed out a systematic
disagreement between melting temperatures of bcc transition metals
measured in shock-waves and DAC experiments. Also in this case,
$T_m$ measured using shock-waves is noticeably higher 
than that obtained by extrapolation of DAC measurements. 
Several possible explanations were proposed to resolve this discrepancy,
including the existence of an extra high P-T phase and an overshoot
of the melting temperature due to the small time scale in
shock-wave experiments~\cite{Errandonea}. We would like to 
suggest that the discrepancy between the melting temperatures determined by these
methods results from different conditions  present at the boundary of the sample.
Inside a DAC, the sample is surrounded by
a pressure transmitting medium. In shock-wave experiments
the molten region inside the sample is bordered by relatively unstressed 
cold regions. In both cases, the sample has no free surface.
It is well known that at  zero pressure,  the mechanism of melting 
differs depending on whether the sample does or does not have  a free surface. 
The purpose of this study
is to examine how the different types of boundary conditions 
systematically affect the melting transition 
at high pressures.

At zero pressure, theories describing the mechanism of melting 
\cite{Dash} can be separated into
two classes. The first one describes the  mechanical melting
of a homogeneous solid resulting from lattice instability
~\cite{Born,Tallon,Wolf,Jin} and/or the spontaneous generation of thermal
defects (vacancies, interstitials, and dislocations)
\cite{Granato,Cahn2,Sorkin,Gomez,Burakovsky,Lund}.
The second class
treats the thermodynamic melting of solids,
which begins at extrinsic defects such as a free surface or an
internal interface (grain boundaries, voids, etc)\cite{Frenken,Phill,Tosati,Barnett,Lutsko,Broughton1,Polturak}. 
From these  studies it is clear that the value of the 
melting temperature is sensitive to 
whether or not the solid has a free surface. The thermodynamic
melting temperature is systematically lower than the melting temperature
of surface-free solids, and the liquid phase always nucleates
on the least closely packed surface.

\begin{figure}
\includegraphics{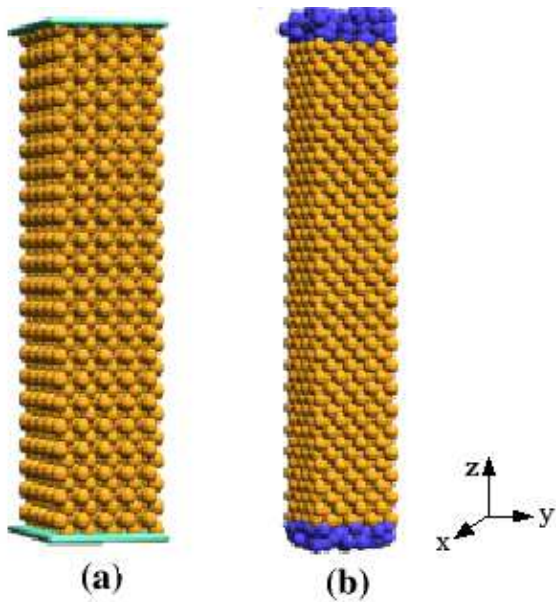}
\caption{\label{hard_wall} (Color online)(a) Snapshot of the Ar(001) sample
bordered by hard walls  at its top and  bottom. (b) Ar(001) sample  bordered by neon layers.
Periodic boundary conditions were applied along x and y. }
\end{figure}

We now examine the question of whether
this distinction affects the interpretation of high pressure experiments.
Here, in order to maintain a high pressure, there can be no free surface.
To shed light on this problem, we decided to simulate melting 
of samples with either ``soft'' or ``rigid'' boundaries.
Specifically, we simulated a system of
argon particles interacting via a pairwise Lennard-Jones (LJ) potential.
In one case the argon was in contact with a rigid wall, represented by
an infinite  step function in the potential. In the second case
we simulated solid argon in contact with a fluid neon layer.
Our model crystal is a  slab made of 44 atomic layers,
and with two surfaces. (See Fig.~\ref{hard_wall}). 
The argon atoms were subjected to periodic boundary conditions only along the
x and y directions (parallel to the free surface).
We studied two different low-index  surfaces:  Ar(011) with 25 
atoms per layer, and  Ar(001) with 32 atoms per layer.
As a reference, we also simulated a surface-free solid sample
with 864 atoms, by applying periodic boundary conditions in all directions.
\begin{figure}
\includegraphics{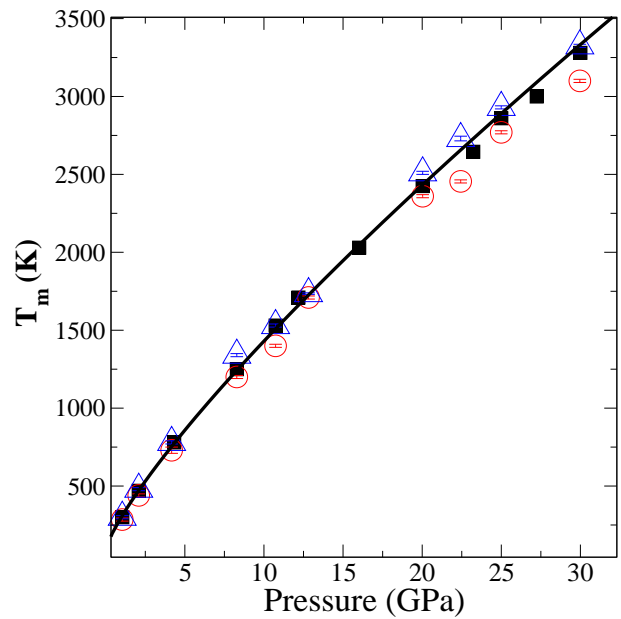}
\caption{\label{pic4}(Color online) Melting temperature as a function of
pressure for the samples with the hard walls: the  Ar(001) sample (triangles, blue online) and
the Ar(011) sample (circles, red online). The solid black squares correspond to the (infinite) surface-free sample.
The solid line is taken from ~\cite{Gomez}. Error bars are smaller than the size of the symbols.}
\end{figure}
In the laboratory, experimental conditions include a fixed pressure P,
temperature T, and  number of atoms N (NPT ensemble).
We performed  Monte Carlo (MC) simulations using this ensemble~\cite{Frenkel}.
The LJ potential was truncated and shifted, with the cutoff distance, $r_c$,
chosen to be $r_c = 2.1\sigma$. The values of the parameters  of the LJ potential
are given in the Table.

The initial conditions in our simulations differed according to the type of
boundary. For the case of hard walls, the distance between the top surface layer of argon and 
the hard wall was set equal to the bulk interlayer distance. In the second case, 
the atoms of neon were initially  arranged in a simple cubic lattice. 
Since the melting temperature, $T_m$, of neon is lower
than that of argon at all pressures, this boundary layer melted immediately
and remained fluid at all temperatures at which simulations were made. 
The interaction between the  Ne-Ne  and Ar-Ne atoms was modeled
using the LJ potential with parameters (see the Table) taken from~\cite{RG}. 
\begin{table}
\caption{\label{tab1} Parameters of the LJ potential}
\begin{tabular}{|c|c|c|} \hline
 Type of atoms & $\epsilon$ (K) & $\sigma$ (A)  \\ \hline \hline
 Ar-Ar  & 0.0104  & 3.4\\ \hline
 Ne-Ne  & 0.0031  & 2.74 \\ \hline
 Ar-Ne  & 0.0061959  & 3.43 \\ \hline
\end{tabular}
\end{table}
Each simulation was started at a low-temperature with a perfect fcc solid sample
at a fixed pressure (P $>$ 1GPa). The temperature of the sample was then gradually raised by
20K - 100K steps, (at low and high pressures respectively) and the sample was equilibrated.
An equilibrium state was considered to be achieved when there was no significant
variation (beyond the statistical fluctuations) of the total energy, pressure, volume and
structure order parameter (the spatial Fourier transform along the [001] direction).
The melting transition was indicated by a jump in the  total energy and volume,
simultaneous with the vanishing of the structure order parameter.
To improve the accuracy in the vicinity of $T_m$, we used  smaller temperature steps of 10 K,
and increased the number of MC steps by a factor of six.  
Throughout this study, interactive visualization (the AViz program~\cite{Adler}) was
implemented to observe sample disorder and melting.
\begin{figure}
\includegraphics{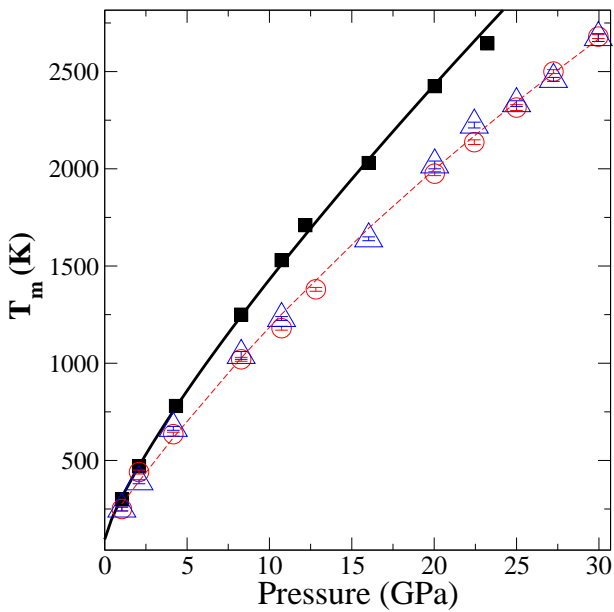}
\caption{\label{pic3} (Color online) Melting temperature as a function of
pressure for the samples with a neon layer at each surface: the  Ar(001) sample (triangles, blue online) and
the Ar(011) sample (circles, red online). The black squares correspond to the (infinite) surface-free sample.
 The dotted line is drawn to guide the eye, and the solid line is taken from ~\cite{Gomez}.
 Error bars are smaller than the size of the symbols. }
\end{figure}
\begin{figure}
\includegraphics{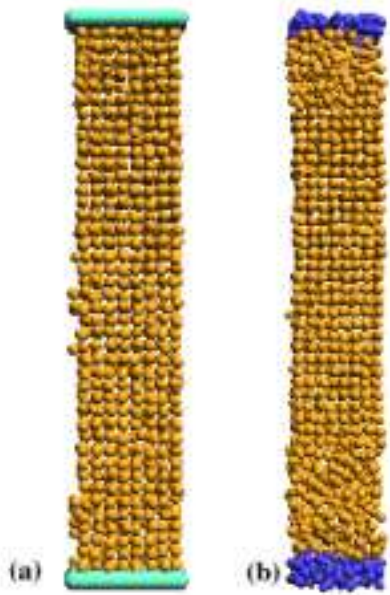}
\caption{\label{ar_ne} (Color online)  Snapshot of the Ar(001) slabs at pressure P = 4.17 GPa:
(a) a sample with hard walls  at T = 740 K ( $T_m$ =  780 K).
(b) a sample with neon layers at T = 625 K ( $T_m$ =  665 K).
Note the presence of premelting near the surface of the Ar-Ne sample. }
\end{figure}
\begin{figure}
\includegraphics{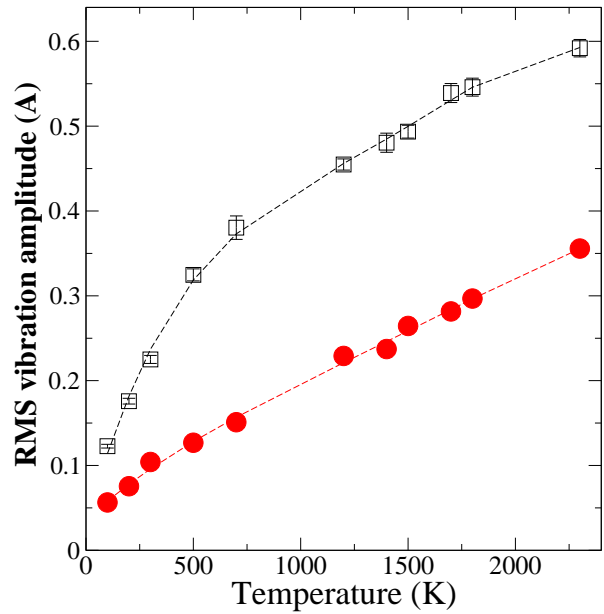}
\caption{\label{u2}(Color online) The depression of the out-of-plane 
atomic vibration amplitude (circles, red online)  relative to the in-plane vibration amplitude (squares),
shown for the (001) sample with  surface atoms bordered by a hard wall.
The pressure is 20 GPa. The dotted lines guide the eye.
 Error bars are smaller than the size of the symbols.}
\end{figure}

The melting curves calculated for
the case of the hard wall are shown in Fig.~\ref{pic4} 
for both the Ar(001) and Ar(011) samples.
For comparison, the points showing $T_m$ of the surface-free solid are also shown.
These points are in very good agreement with a simulation (solid curve in
Fig.~\ref{pic4}) of a surface-free
solid made by Gomez et. al.~\cite{Gomez}. 
It is seen that the argon sample  bordered by hard walls melted
at a temperature very close to that of a surface-free solid. 
The sample with the (011) surface melted at a slightly lower temperature
than the sample with the (001) surface.

The melting curves calculated for
the  Ar bordered by fluid neon are shown in  Fig~\ref{pic3}
for the Ar(001) and Ar(011) samples. The curves are compared
with that  calculated for the surface-free solid.
Within our resolution we did not observe a difference in $T_m$ between
samples with the (001) and the (011) surface.

A comparison of the melting curves for the samples
with soft and rigid boundaries shows that the  sample bordered by the neon layer melted at a 
systematically lower temperature than the sample with the hard walls.
Another important difference, shown in  Fig.~\ref{ar_ne}, is that
premelting effects were absent in the  sample with the hard walls, whereas
in the case of the neon covered surface a gradual premelting was observed.

We interpret the above results as follows:
the interactions with the hard wall seem to 
effectively inhibit the out-of-plane motion of the surface atoms.
This result is shown in Fig.~\ref{u2}. In contrast, the 
in-plane  and out-of-plane RMS vibration amplitude in the sample bordered
by fluid neon is approximately  the same. 
Restriction of the out-of-plane motion suppresses
thermal disordering of the surface.
Absence of thermal disordering inhibits surface premelting
and allows superheating up to the temperature at which crystal
lattice becomes unstable.
Consequently, superheating of argon bordered by hard walls
is possible. The situation is analogous 
to the well-known experiment by Daeges et. al.~\cite{Daeges}
in which superheating of silver coated with gold
was demonstrated (gold has a higher $T_m$ than silver).

In our opinion, the conditions in the simulations with the hard walls
are similar to those found in the shock wave experiments.
The simulations can be related to the experiments in the following way: 
Typically, the part of the solid  which is compressed during the 
propagation of the shock wave is much smaller than the size of the sample.
Therefore, the instability occurs inside a region surrounded by
a relatively cold material, which can act as a hard wall. Further
support for this conjecture comes from the work of Kanel et. al.~\cite{Kanel}
who clearly observed superheating of aluminum single crystals with the shock
wave technique.
Another example where superheating is distinctly observed
is in the case of compressed argon bubbles inside an aluminum matrix~\cite{Rossouw}.
In this experiment the free surface of the solid was eliminated 
and as a result the solid was superheated.
Therefore, the melting transition is closer
to mechanical melting triggered by lattice instability.

On the other hand, in DAC experiments the material under study
is usually surrounded by a hydraulic medium~\cite{Errandonea}
so that it surface is in contact with a rare-gas or some other
inert material. In addition the heating is usually done by a laser
which heats mainly the surface. This situation
is close to our simulations with the fluid neon layer.
According to the results of the simulations,
melting in this case is more like thermodynamic melting.

Before concluding we remark that the LJ (6,12) potential
is not accurate enough at high pressures to allow
quantitative comparison with experiment~\cite{Datchi,Boehler,Belonoshko0}. 
However, we believe
that the  generic nature of our results is valid.

In conclusion, we simulated the melting of a solid in
the presence of two types of pressure transmitting medium at the sample boundaries.
We found that with the soft medium (liquid neon layers) melting is closer to thermodynamic,
nucleating at the surface, while with the rigid medium (hard walls) 
the solid exhibits  superheating and melts via a lattice instability.
These results are related to high pressure melting experiments
and appear to be consistent with  systematic
differences that exist between shock wave and DAC measurements.   
We believe that the  disparities between the results of measurements obtained
with these two techniques at least to some degree originate
in the different conditions at the solid-liquid interface.
We suggest that  results obtained with a DAC technique  should
be compared with thermodynamic theories, while shock wave results
should be compared with theories based on a mechanical instability.

This study
was supported in part by the Israel Science Foundation and by the
Technion VPR fund for the promotion of research.

\end{document}